# Software Effort Estimation with Ridge Regression and Evolutionary Attribute Selection

Efi Papatheocharous[1], Harris Papadopoulos[2] and Andreas S. Andreou[3],

[1] University of Cyprus, Department of Computer Science. 75 Kallipoleos Street,
P.O. Box 2053, CY1678 Nicosia, Cyprus
efi.papatheocharous@cs.ucy.ac.cy
[2] Frederick University, Computer Science and Engineering Department.
7 Y. Frederickou Str., Palouriotisa, 1036 Nicosia, Cyprus
h.papadopoulos@frederick.ac.cy
[3] Cyprus University of Technology, Department of Electrical Engineering and Information Technology. 31 Archbishop Kyprianos Street, 3036 Lemesos, Cyprus
andreas.andreou@cut.ac.cy

**Abstract.** Software cost estimation is one of the prerequisite managerial activities carried out at the software development initiation stages and also repeated throughout the whole software life-cycle so that amendments to the total cost are made. In software cost estimation typically, a selection of project attributes is employed to produce effort estimations of the expected human resources to deliver a software product. However, choosing the appropriate project cost drivers in each case requires a lot of experience and knowledge on behalf of the project manager which can only be obtained through years of software engineering practice. A number of studies indicate that popular methods applied in the literature for software cost estimation, such as linear regression, are not robust enough and do not yield accurate predictions. Recently the dual variables Ridge Regression (RR) technique has been used for effort estimation yielding promising results. In this work we show that results may be further improved if an AI method is used to automatically select appropriate project cost drivers (inputs) for the technique. We propose a hybrid approach combining RR with a Genetic Algorithm, the latter evolving the subset of attributes for approximating effort more accurately. The proposed hybrid cost model has been applied on a widely known high-dimensional dataset of software project samples and the results obtained show that accuracy may be increased if redundant attributes are eliminated.

## 1 Introduction

Software companies and other stakeholders, such as, customers, end-users, managers and researchers, have worked over the past decades on improving the accuracy and consistency of effort estimations for developing software systems. To achieve this, several techniques have been developed aiming on one hand to develop high quality software with the lowest possible costs and on the other hand to strive towards successful project completion [1]. Moreover, developing techniques to efficiently estimate the overall development costs, especially from the planning phases of a project, can offer a significant competitive advantage for software companies and project managers. This advantage can be used for reducing the risk of resource misallocation and enhancing the manager's ability to deal with budgeting, planning and controlling of the project processes and assets.

Software cost models proposed in literature are based on expert judgement or employ some mathematical or machine learning technique to reach to improved effort approximations. Models usually employ historical project data obtained from previous software developments and very rarely the suitability of this data is questioned. Moreover, for a software cost model to be regarded as practical the following requisites need to co-exist; firstly, the delivery of accurate, transparent and meaningful effort approximations and secondly, the utilisation of suitable, measurable and available information at the project initiation stages, where effort estimation is needed the most. Nevertheless, the majority of cost models found in the literature require project information that is available only after requirements specification and relate for instance to the size of the software, application type, function points and language type used.

Experienced cost estimators or project managers need to manually perform evaluation, filtering and identification of the appropriate project parameters that should be used in each model as a pre-processing step. This task will be subsequently combined with obtaining measurements from real-life projects, evaluating the measurements through cost models and maintaining them for future estimations. Carrying out these processes manually is labour intensive and thus, focusing only on measuring and maintaining a subset of relevant project attributes that will direct cost models to better effort approximations, could lead in reducing a lot of time and effort required by project managers.

Our previous work has targeted the production of reliable confidence measures based on the dual form Ridge Regression (RR) technique [2] where it was suggested that results could be improved if the most relevant project attributes were selected and engaged in the prediction process. RR has also been successfully applied in the past in the area of software engineering by various researchers yielding promising results. Specifically, RR has been used to estimate the coefficients for the COCOMO model [3], one of the most widely known cost models, to produce classification scores and remove unnecessary features [4] and to improve the performance of regressions on multi-collinear datasets [5]. However, none of these works explored the identification of the optimum feature set for estimating effort.

In this paper, we propose the use of a Genetic Algorithm (GA) for automatically identifying the optimal subset of cost drivers participating in the cost estimation process of the dual variables RR technique. The advantage of the dual variables version of the technique is that it uses the so-called kernel trick to efficiently construct non-linear regressions. The proposed approach is tested on a real world dataset and the results indicate that the selection of appropriate attributes can not only reduce the information required to approximate effort, but it can also result in obtaining better approximations.

The rest of this paper is organised as follows: Section 2 gives an overview of related work on the problem of software cost estimation. It also discusses common difficulties of data-driven cost estimation models and presents the attempt of several researchers to eliminate the number of features used in cost estimation models. Section 3 describes the dual form of Ridge Regression (RR) algorithm. Section 4 presents the Genetic Algorithm (GA) for evolving the selection of attributes. Section 5 gives a detailed description of the experiments conducted and discusses the main results obtained. Finally, Section 6 summarises our conclusions and future research plans.

## 2 Related Work

Data-driven models are considered the most popular approaches in software cost estimation. They employ historical project data for building and calibrating mathematical formulas that relate cost drivers with effort. These models require usually a wealth of data and make use of numerical transformations to achieve accurate and reliable effort estimates [6]. The most common parametric estimation models, such as the COCOMO, use linear regression to approximate effort [7]. Naturally, the accuracy of data-driven models depends mainly on the attributes selected and included in the estimate. In addition, data-driven methods frequently need to handle project data that are not only of numerical type, but also of binary, categorical and nominal type. Therefore in some cases, they need to be extended to handle transformations and to offer capabilities closer to reasoning and deduction. Thus, searching for mathematical models that capture relationships between the contributing factors and effort usually results in constructing considerably complex formulas.

Related studies questioned the necessity of a large number of features involved usually in cost estimation techniques and some investigations showed that in most cases redundant features could be eliminated. In [8] the accuracy of the COCOMO was improved when a wrapper technique was used to identify the most promising features for the model. Other Feature Subset Selection (FSS) algorithms were investigated for increasing the accuracy performance in analogy-based software effort estimation models [9]. In [10] Genetic Algorithms (GA) were used to assign proper weights to features, and three different heuristics were proposed to increase the estimation performance. The authors identified that even though their approach presents certain advantages, GAs are random, greedy and iterative methods that are always prone to get stuck in local minima.

The aforementioned difficulties led to a number of studies investigating further the cost estimation issue with computational intelligent methods, such as artificial neural networks [11, 12], fuzzy logic modelling [13, 14] and evolutionary algorithms [15, 16] and seeking for optimised effort approximations. The popularity of research on alternative models has increased, according to systematic reviews [7], even though the obtained results are always compared to those of linear regression techniques as a baseline. Moreover, many of these models have been extended into hybrid forms in an attempt to improve their intuitiveness, accuracy and robustness in a number of studies e.g., [7, 17].

As previously mentioned, the largest portion of research studies in cost estimation are using linear regression as the most applicable method which also serves as benchmark for assessing performance in terms of accuracy [7]. Recently dual variables RR has been proven promising for driving accuracy performance to higher levels [2, 5]. The authors' focus in [2] was on deriving confidence intervals for effort estimation and in [5] it was on the multi-collinearities of datasets which were found to lead to unstable regression coefficients. This work studies the genetic evolution of cost attribute subsets for RR experimenting on a widely known, large and real life dataset. The aim is to both improve results and simplify the resulting model. The dataset selected for experimentation has a multi-dimensional form, contains a plethora of software projects and a large number of categorical and multi-valued attributes. The choice of the particular dataset was made due to its high complexity and big size.

## 3  Dual Variables Ridge Regression

Ridge Regression (RR) is an improvement of the classical Least Squares technique which is one of the dominant methods applied in software cost estimation and one of the most widely used regression algorithms in a large number of research areas such as ionospheric parameter prediction [18], ecological risk assessment [19] and face recognition [20]. In this work we use the dual variables RR method, proposed in [21], as it employs kernel functions to allow the construction of non-linear regressions without having to carry out expensive computations in a high dimensional feature space.

To approximate a set of sample projects $\{(x_1,y_1), \ldots, (x_l,y_l)\}$, where $x_i \in \Re^n$ is the vector of attributes (cost drivers) for project $i$ and $y_i \in \Re$ is the effort of that project, the well known RR procedure recommends finding the $w$ which minimises

$$a\|w\|^2 + \sum_{i=1}^{l}(y_i - w \cdot x_i)^2, \tag{1}$$

where $a$ is a positive constant, called the *ridge parameter*. Notice that RR includes Least Squares as a special case (by setting $a = 0$). The RR prediction $\hat{y}_t$ for an input vector $x_t$ is then $\hat{y}_t = w \cdot x_t$.

The dual variables formula, derived in [21], for the prediction of an input vector $x_t$ is

$$\hat{y}_t = Y(K + aI)^{-1}k \tag{2}$$

where $Y = (y_1, \ldots, y_l)$ is the vector consisting of the effort outputs of the projects in the training set (the set of known project samples), $K$ is the $l \times l$ matrix of dot products of the input vectors $x_1, \ldots, x_l$ of those data samples,

$$K_{i,j} = K(x_i, x_j), \quad i = 1,\ldots,l, \quad j = 1,\ldots,l \tag{3}$$

$k$ is the vector of dot products between $x_t$ and the input vectors of the training samples,

$$k_i = K(x_i, x_t), \quad i = 1,...,l, \tag{4}$$

and $K(x, x')$ is the kernel function, which returns the dot product of the vectors $x$ and $x'$ in some feature space.

In this work we used the RBF kernel function, which is the typical choice of kernel in machine learning literature and is defined as

$$K(x, x') = \exp(-\frac{\|x - x'\|^2}{2\gamma^2}). \tag{5}$$

## 4 Genetically Evolved Cost Driver Subsets

A Genetic Algorithm (GA) is an optimisation technique inspired by natural evolution. It evolves a population of encoded solutions (called individuals) to a given problem through generations using genetic operations. The typical genetic operations used are crossover, or recombination, and mutation. The individuals at each step, called a generation, are evaluated and a score is assigned to each one, called its fitness value, indicating how good the solution it represents is. The fitness of each individual defines its likelihood of being selected for the next generation. Until a new generation is completed, individuals from the current generation are selected probabilistically based on their fitness to generate offspring for the new generation. There are also a few individuals, the fittest ones, which are carried forward to the new generation unchanged, that is, without the application of any genetic operation on them. The same process is repeated until an optimal solution is reached or a stopping criterion is met, which in many cases is a maximum number of generations. The solution represented by the fittest individual in the last population is the one adopted as the resulting solution of the GA.

In our case each individual is a bit string of the size of the cost drivers contained in the dataset. The cost drivers represented by the bits set to 1 are taken into account as inputs while all others (set to 0) are not. To evaluate each individual we perform a 10-fold cross-validation process on the set of known examples (the training set) using as inputs only the selected cost drivers. Specifically, we split the training set into 10 parts of almost equal size and we apply the RR technique (with only the selected cost drivers) 10 times, each time evaluating its performance on one part after training on the other nine. For the performance evaluation we calculate the *Magnitude of Relative Error* (*MRE*) of each project $i$ as:

$$MRE_i = \left|\frac{y_i - \hat{y}_i}{y_i}\right|. \tag{6}$$

At the end of the 10-fold cross-validation process the *Mean Magnitude of Relative Error* (*MMRE*) over the whole set is calculated as the mean value of the *MRE*s of all projects. Since the smaller the *MMRE* the fitter the corresponding individual, the fitness of each individual is defined as:

$$\text{fitness} = -MMRE. \tag{7}$$

To create each generation we first select the 10% fittest individuals and place them in the new generation as they are. Then until the new generation is complete we select two individuals at a time, which we recombine and mutate to generate two new individuals for the new population. To select the individuals we perform stochastic uniform selection based on the rank of the individuals in the current population. In effect this selection function lays out a line in which each individual corresponds to a section of the line of length proportional to its rank. The function then moves along this line in steps of equal size selecting the individual from the section it lands on. The size of the first step is a uniform random number less than the fixed step size.

The crossover function we use is the uniform crossover, with a probability of being applied to each pair, called crossover rate, of 0.8. This function creates a random bit string of the same size as the two parents, called crossover mask, and generates the first child by copying the parts of the first parent at the points where the crossover mask has a 1 and the parts of the second parent at the points where the crossover mask has a 0. For the second child the same process is repeated with the parents reversed. We also use uniform mutation, which flips each bit of an individual with a given probability, which in this case was set to 0.01.

## 5 Experiments and Results

The proposed genetically evolved algorithm of software cost drivers was implemented using the MATLAB (R2008b) Genetic Algorithm and Direct Search Toolbox [22] and had a two-fold aim: Firstly, select the best subset of cost drivers from a wider pool of drivers, and secondly minimise the overall relative error rate of the RR technique.

The ISBSG dataset used in the experiments [23] includes a vast number of project attributes related to the application domain, programming language used, language type, development technique, resource level, functional size of the software produced, etc. The dataset contained numeric, categorical, multi-categorical and null values and so, we performed the following actions of column and row filtering:
  a) Excluded attributes that have been measured after the project completion and would not be therefore practical for the cost estimation model we were building.
  b) Calculated the number of null records for the remaining attributes and excluded those that presented more than 40% because they would cause dramatic decrease to the useful sample finally used.
  c) Referred to the quality rating provided by the ISBSG (the organization from which the data used in this study is drawn) reviewers and excluded project information that were rated lower to grade 'B'.
  d) Excluded records that contained null values in numerical attributes.
  e) Created new binary columns for different values of categorical attributes. In these columns and for each project reported the value of 1 if it belonged to the new categories created, or 0 if it did not belong.

f) All numerical values were normalized to the minimum and maximum range of 0 and 1 respectively so that they all had the same impact.

From the original software attributes reported within the dataset, we included in our experiments only those that contained enough data to be useful and meaningful for cost estimation and the final ISBSG dataset after this preprocessing contained 467 projects and 82 attributes. Therefore, the bit string representation used for the individuals of the algorithm had 82 bits and all combinations of selected attributes were treated as solutions to the problem.

### 5.2 Results

The summary statistics of the output (dependent) effort variable of the ISBSG sample used in the experiments is shown in Table 1. The figures show major differences between the actual data used for the estimation.

Table 1: Summary statistics of dependent variable (full-cycle summary work effort)

| **Mean** | **Std. Error** | **Median** | **Mode** | **Std. Dev.** |
|---|---|---|---|---|
| 6,644.43 | 595.62 | 2,385.00 | 1,238.00 | 12,871.37 |
| **Sample Variance** | **Kurtosis** | **Skewness** | **Min** | **Max** |
| 165,672,215.32 | 38.86 | 5.08 | 97.00 | 150,040.00 |

Table 2: Results for the ISBSG dataset - Genetically evolved cost drivers for RR

| Partition | Training Set | | | Testing Set | |
|---|---|---|---|---|---|
| | *MMRE* full | *MMRE* selection | No. of selections | *MMRE* full | *MMRE* selection |
| 1 | 0.314 | 0.374 | 33 | 0.763 | 0.733 |
| 2 | 0.335 | 0.418 | 30 | 0.524 | 0.490 |
| 3 | 0.333 | 0.415 | 29 | 0.506 | 0.470 |
| 4 | 0.328 | 0.397 | 34 | 0.502 | 0.521 |
| 5 | 0.355 | 0.424 | 35 | 0.434 | 0.445 |
| 6 | 0.312 | 0.374 | 30 | 0.815 | 0.748 |
| 7 | 0.319 | 0.398 | 32 | 0.551 | 0.538 |
| 8 | 0.328 | 0.375 | 40 | 0.549 | 0.513 |
| 9 | 0.337 | 0.411 | 32 | 0.657 | 0.745 |
| 10 | 0.344 | 0.421 | 33 | 0.547 | 0.495 |

In order to test the performance of the proposed approach we randomly partitioned the dataset into training and testing sets 10 times, each time allocating 80% of the projects to the training set and the remaining 20% to the testing set. Our GA was then applied to each of the resulting training sets with a population of 100 individuals for 100 generations producing the evolved subset of selected cost drivers for the corresponding partition. The selected cost drivers were then used as inputs for training the RR technique on the training set of each partition and evaluating its cost estimates on both the training and testing sets of the partition.

The obtained results are summarised in Table 2. The first column of this table indicates the sequence number of the dataset partition. The second and fifth columns report the error figures obtained when using the complete set of inputs for training and evaluating the algorithm while the third and sixth columns report the corresponding error figures obtained when using only the selected subset of inputs. Finally the fourth column indicates the number of attributes included in the selected subset of the GA.

The values reported in the second and third column of Table 2 show that the performance of RR on its training sets deteriorates slightly with the reduction of the used attributes. This is of course natural since by reducing the number of inputs we also reduce the number of free parameters and therefore the ability of the function to fit the training data. As a result, by removing inputs which are either misleading or contain little information relevant to our problem, we in effect reduce the degree of overfitting that occurs.

The actual improvement resulting from the use of only the selected attributes is evident from the performance of RR on the testing sets reported in the fifth and sixth columns of the table. Here we see that the selection of attributes made by the GA reduced the resulting error in 7 out of 10 cases. The mean values over the 10 partitions give an *MMRE* of 0.5848 when using all attributes compared to 0.5698 when using only the selected attributes. Furthermore, this is achieved by selecting always less than half of the available cost drivers and on average only 33 out of the full 82. This improvement in *MMRE* values on the testing set also confirms that the increase of *MMRE* observed on the training set represents a reduction of overfitting.

Finally, it would be very interesting to analyse the actual attributes selected by the GA, but due to the large original size of the dataset (82 attributes), this is not practical and requires further investigation, filtering or weighting. However, our results have shown that the proposed approach can be used successfully for identifying the appropriate attributes to measure and maintain for future cost estimations and therefore saving an important amount of time and effort that is spent in acquiring and maintaining large quantities of data.

## 6   Conclusions

This paper proposed the use of the dual variables Ridge Regression (RR) technique in conjunction with a Genetic Algorithm (GA) to evolve the selection of cost drivers used as inputs for the estimation of software effort. An important advantage of using a GA is that it can search the vast space of possible combinations of cost drivers efficiently and reach a near-to-optimal outcome. The results obtained from the evaluation phase of our experiments show that the proposed algorithm did not only reduce the number of cost drivers used to less than half, but it also resulted in an improvement to the performance of the dual variables RR technique.

In the future we plan to further investigate the use of our approach in combination with other techniques such as stepwise regression, greedy hill climbing, simulated annealing and artificial neural networks. We also plan to study the effect that each cost driver has on the resulting predictions, both for the ISBSG dataset used in this work and for other datasets. The results of such a process may enable decision makers, project managers or cost estimators to make more informed decisions about the project parameters that should be measured and used for producing accurate effort estimates. Another future direction of this work is to extend the Genetic Algorithm developed into including the optimisation of the Ridge Regression parameter values, i.e., the ridge factor and RBF spread, as part of the implementation, so that the model becomes properly calibrated for the selected cost drivers.